# Probing Magnetic Configurations in Co/Cu Multilayered Nanowires

Jared Wong, Peter Greene, Randy K. Dumas, Kai Liu*

*Physics Department, University of California, Davis, CA  95616*

## Abstract

Magnetic configurations in heterostructures are often difficult to probe when the magnetic entities are buried inside. In this study we have captured magnetic and magnetoresistance "fingerprints" of Co nanodiscs embedded in Co/Cu multilayered nanowires using a first-order reversal curve method. In 200nm diameter nanowires, the magnetic configurations can be tuned by adjusting the Co nanodisc aspect ratio. Nanowires with the thinnest Co nanodiscs exhibit single domain behavior, while those with thicker Co reverse via vortex states. A superposition of giant and anisotropic magnetoresistance is observed, which corresponds to the different magnetic configurations of the Co nanodiscs.

Multilayered magnetic nanowires have been a model system for heterostructured junctions that exhibit a host of fascinating perpendicular spin transport phenomena, such as current-perpendicular-to-plane giant magnetoresistance (GMR),[1-3] tunneling magnetoresistance[4] and spin-transfer torque effects.[5] Due to the extremely small physical dimensions the magnetic components in these nanowires or junctions often exhibit vortex state during magnetization reversal.[6-10] The vortices can drastically affect the spin-dependent transport properties,[11] and even offer opportunities for data storage based on the vortex core states.[12] Local detection of vortex state in magnetic nanostructures has often been performed by magnetic imaging techniques.[7, 8] However, this is generally challenging in hybrid structures where the magnetic elements are underneath other surfaces or embedded inside a growth matrix. On the other hand, conventional magnetometry techniques can measure buried magnetic elements, even those with dimensions below the magnetic imaging resolution. But for arrays of nanowires or junctions they suffer from the array averaging effect where distinct features of the hysteresis loop may be washed out due to property variations.

Recently, we have used a magnetometry-based first-order reversal curve (FORC) technique to investigate sub-100 nm Fe nanodots.[13, 14] The irreversible vortex nucleation and annihilation events are used to track the nanomagnets that reverse via a vortex state. In this study, we investigate the magnetization reversal in Co/Cu multilayered nanowires where the Co segments are nanodiscs embedded inside the wires. The aspect ratio of the Co nanodiscs is tuned to adjust the magnetic configurations. The magnetic and magnetoresistance (MR) characteristics of the nanowires are captured.

The nanowires are grown by pulsed electrochemical deposition into nanoporous polycarbonate membranes. The membranes are 6-10 μm in thickness, with pore diameters of 50-200 nm and a pore density of 3-6 $\times 10^8$ pores/cm$^2$. A 360 nm layer of copper is sputtered onto one side of the membrane and serves as the working electrode. The electrolyte and deposition conditions have been reported in prior studies.[3] Relative to the Ag$^+$/AgCl reference electrode, Cu and Co are deposited at -0.4 V and -0.9 V, respectively, using a Princeton Applied Research Potentiostat 263A. The nanowires are capped by another 350 nm layer of Cu sputtered on the other side of the membrane.

The deposited nanowires have been studied by x-ray diffraction, which reveals a polycrystalline structure. Free-standing nanowires have been obtained by etching away the polycarbonate membrane. Scanning and transmission electron microscopy have been used to characterize the nanowires. Composition variation of the Co/Cu multilayered nanowires has been verified using energy dispersive x-ray microanalysis scan along the wire.

Magnetic properties of the nanowires have been measured by a PMC hybrid vibrating sample and alternating gradient magnetometer with nanowires inside the membrane. The FORC technique has been employed to study details of the magnetization reversal. After saturation, the samples are brought to successively more negative reversal fields $H_R$ and magnetization $M$ is measured at increasing applied field $H$ to trace out FORC's, following prior procedures.[15, 16] The FORC distribution $\rho \equiv -\partial^2 M(H, H_R)/2\partial H \partial H_R$, which captures the irreversible magnetization switching, is plotted against $(H, H_R)$ coordinates on a contour map. Alternatively • can be plotted in $(H_C, H_B)$ through a simple rotation of the coordinate system defined by: $H_B=(H+H_R)/2$ and $H_C=(H-H_R)/2$, where $H_C$ is the local coercive field and $H_B$ is the local interaction or bias field.

Electrical resistance and MR have been measured on a selected collection of nanowires. Two orthogonal thin strips of silver paint (<0.5 mm) are applied on both sides of the Cu-coated membranes. The unmasked Cu is etched away, leaving behind a small area in between the crossed strips with connected nanowires. Current and voltage leads are then attached to the strips using silver epoxy to allow electrical measurement.

Representative room temperature magnetic major hysteresis loops of 200 nm diameter Co/Cu multilayered nanowires are shown in Fig. 1. The Co segments have variable thickness in the range of 10-55 nm, separated by 250-350 nm thick Cu pillars so that the dipolar interactions are minimized.[17, 18] Since these multilayered nanowires are essentially collections of Co nanodiscs, the shape anisotropy forces the magnetic easy axis to be in-plane, i.e., perpendicular to the wires. *In the following we concentrate on this field orientation.* The coercivity of [Co(10nm)/Cu(250nm)]$_5$ nanowires is 460 Oe and the normalized remanent magnetization $M_r/M_S$ is 46 % [Fig. 1(a)]. In [Co(55nm)/Cu(350nm)]$_4$ nanowires, the coercivity and remanence are reduced to 170 Oe and 13%, respectively [Fig. 1(b)]. Also, the hysteresis loop develops a slight pinch near zero field. We have examined the width of the loop $W$ ($M/M_S$) at various $M/M_S$ values and compared with twice the coercivity $W$ ($M/M_S$=0) [Fig. 1(c)]. At small Co layer thickness (10 and 32nm), the width analysis yields a single parabola, with a maximum value of $W$ ($M/M_S$) / $W$ (0) = 1 at $M/M_S$ =0, indicating that the widest part of the hysteresis loop intersects the field axis. As Co thickness reaches 37 nm and beyond, a double peak characteristic is observed. The wing-like pattern indicates that the widest parts of the loop are in the first and third quadrants, resulting in a pinched loop. This pattern is often characteristic of reversal via a vortex state, which ideally has zero remanence due to the flux closure state. However, as shown before, the loop shape alone is not a reliable indicator of the reversal mechanism.[19]

To conclusively distinguish the reversal mechanisms we have employed the FORC method, which has been shown to be very sensitive to irreversible magnetization switching, such as the vortex nucleation and annihilation events.[13, 14] For the 200 nm diameter [Co(10nm)/Cu(250nm)]$_5$ nanowires with the thinnest Co nanodiscs, Fig. 2(a) shows a family of FORC's measured at room temperature. The corresponding FORC distribution $\rho$ is shown in Fig. 2(b), exhibiting a prominent ridge along the local coercivity $H_C$ axis, centered at ($H_C$ = 0.6 kOe, $H_B$ =0). This FORC distribution is characteristic of single domain nanodots[13, 14] and nanoparticles.[20] The limited spread along the $H_B$ axis is a manifestation of little interactions amongst the Co nanodiscs, due to the separation by the relatively thick Cu spacers.[18] The spread along the local coercivity is a result of the variations in Co nanodiscs inside the nanowires.

In contrast, the 200 nm diameter [Co(55nm)/Cu(350nm)]$_4$ nanowires with thicker Co exhibit completely different FORC characteristics. The FORC's fill the interior of the major loop rather unevenly [Fig. 2(c)]. Near zero field, the FORC's vary linearly with magnetic field, indicating reversible processes. The FORC distribution in $H$–$H_R$ coordinates is shown in Fig. 2(d), which is a rotation of the $H_B$-$H_C$ coordinates. It displays a complex butterfly-like pattern with three prominent features. This pattern is strikingly similar to that seen in Fe nanodots and indicates that the magnetization reverses via a vortex state.[13, 14] It illustrates irreversible switching mainly along two FORC's, with $H_R$ ~ 0.1 kOe and -1.4 kOe, respectively. The first feature centered at ($H$, $H_R$) = (0.4, 0.1) kOe corresponds to the annihilation of the vortices: as the reversal field $H_R$ is reduced from positive saturation, vortices are nucleated at 0.1 kOe; as the applied field $H$ is increased, the vortices are annihilated at 0.4 kOe from the same side of the Co nanodiscs that they have nucleated from. Similarly the other two features centered at ($H$, $H_R$) = (-0.1, -1.4) kOe and (1.4, -1.4) kOe correspond to the nucleation of vortices from the opposite side of the Co nanodiscs and the subsequent annihilation events, respectively.

The characteristic FORC distributions show that in 200 nm diameter [Co(10nm)/Cu(250nm)]$_5$ nanowires single domain state dominates the Co reversal; whereas in [Co(55nm) / Cu(350nm)]$_4$ nanowires vortex state dominates. Indeed this has been conformed by micromagnetic simulations on 200 nm diameter Co nanodiscs with 10 nm and 50 nm thickness using the OOMMF code.[21]

Magnetoresistance of the Co/Cu multilayered nanowires have been measured at room temperature. In a test sample of 50 nm diameter [Co(5nm)/Cu(8nm)]$_{400}$ nanowires a 10% GMR, defined as [R(0)-R(H)]/R(H), is observed for both field parallel and perpendicular to the wire geometries. The result is consistent with previous studies on such Co/Cu nanowires.[3] To test the magnetic configurations in 200 nm diameter nanowires, MR is measured for a [Co(50nm)/Cu(10nm)]$_{150}$ sample, as shown in Fig. 3. The Co thickness is in the range where vortex state is expected in individual Co nanodiscs. When the field is applied parallel to the wires, a double peak feature is observed [Fig. 3(a)]: along the decreasing field sweep from 10 kOe, a peak appears at a negative field of -0.5 kOe; along the increasing field sweep, a symmetric peak appears at +0.5 kOe. Interestingly, when the field is applied perpendicular to the wires, a 4-peak pattern is observed [Fig. 3(b)]. For example, along the decreasing field sweep from 10 kOe, MR first reaches a maximum at a *positive field* of +0.5 kOe, followed by a second smaller peak at a negative field of -0.8 kOe.

The MR features are due to the superposition of GMR and anisotropic MR (AMR), the latter of which is described by:[22]

$$R(\theta) = R_T + \Delta R \cos^2 \theta \tag{1}$$

where resistance $R$ depends on the angle $\theta$ between the applied field and current direction, and $R_T$ is the resistance in the transverse direction. The longitudinal resistance $R_{//} = R_T + \Delta R$ when $\theta = 0$. In our measurement, when the field is parallel to the wires, the MR is primarily the GMR effect due to spin-dependent scattering, leading to the typical double peak pattern. The peak positions correspond to states with maximal spin disorder near the coercive fields. Therefore along the decreasing field sweep the GMR peak occurs in a *negative* field and *vice versa* for the increasing-field sweep. When the field is perpendicular to the wires (thus parallel to the Co nanodiscs), the GMR contributions still lead to the two smaller peaks seen in Fig. 3(b). Additionally, AMR leads to the other two bigger peaks, consistent with reversal via a vortex state: along the decreasing field sweep, nucleation of a vortex core occurs at a *positive field*, as confirmed by the aforementioned positive nucleation field captured by FORC; this introduces magnetic moments out of the Co nanodisc plane; consequently, the AMR contribution of this vortex core results in an increase of resistance from $R_T$ to $R_{//}$ and leads to the observed MR maximum. Note that in this [Co(50nm)/Cu(10nm)]$_{150}$ nanowire sample, the Cu spacer is much thinner and the interlayer coupling between adjacent Co nanodiscs is not negligible[18] and likely helps to stabilize the vortex structure during the magnetization reversal. A potential extreme case is that the adjacent Co segments couple strongly together to behave like continuous Co nanowires; in that case when the field is perpendicular to wires the AMR should have double positive peaks [similar to Fig. 3(a)], which is not what we have observed in Fig. 3(b).

In summary, we have investigated the magnetization reversal mechanisms in Co nanodiscs embedded in Co/Cu multilayered nanowires by changing the Co segment aspect ratio. In 200nm diameter Co/Cu nanowires where Co nanodiscs are well separated by Cu spacers, single domain behavior is dominant in Co nanodiscs with a thickness of 10-32nm. When the Co thickness increases to 37-55nm, vortex state reversal dominates. Magnetoresistance results of the nanowires illustrate contributions from both GMR and AMR, consistent with the formation of a vortex core during reversal.

This work has been supported in part by CITRIS, NSF (ECCS-0725902 and PHY-0649297) and the Alfred P. Sloan Foundation. We thank Daniel Masiel and Nigel D. Browning for  electron microscopy studies of the nanowires.

**REFERENCES**

\*    Electronic mail: kailiu@ucdavis.edu

[1]   L. Piraux, J. M. George, J. F. Despres, C. Leroy, E. Ferain, R. Legras, K. Ounadjela, and A. Fert, Appl. Phys. Lett. **65**, 2484 (1994).

[2]   A. Blondel, J. P. Meier, B. Doudin, and J. P. Ansermet, Appl. Phys. Lett. **65**, 3019 (1994).

[3]   K. Liu, K. Nagodawithana, P. C. Searson, and C. L. Chien, Phys. Rev. B **51**, 7381 (1995).

[4]   B. Doudin, G. Redmond, S. E. Gilbert, and J. P. Ansermet, Phys. Rev. Lett. **79**, 933 (1997).

[5]   L. Piraux, K. Renard, R. Guillemet, S. Matafi-Tempfli, M. Matefi-Tempfli, V. A. Antohe, S. Fusil, K. Bouzehouane, and V. Cros, Nano Lett. **7**, 2563 (2007).

[6]   R. Cowburn, D. Koltsov, A. Adeyeye, M. Welland, and D. Tricker, Phys. Rev. Lett. **83**, 1042 (1999).

[7]   T. Shinjo, T. Okuno, R. Hassdorf, K. Shigeto, and T. Ono, Science **289**, 930 (2000).

[8]   A. Wachowiak, J. Wiebe, M. Bode, O. Pietzsch, M. Morgenstern, and R. Wiesendanger, Science **298**, 577 (2002).

[9]   K. Guslienko, V. Novosad, Y. Otani, H. Shima, and K. Fukamichi, Phys. Rev. B **65**, 024414 (2002).

[10]  H. F. Ding, A. K. Schmid, D. Li, K. Y. Guslienko, and S. D. Bader, Phys. Rev. Lett. **94**, 157202 (2005).

[11]  J. Shi, S. Tehrani, and M. R. Scheinfein, Appl. Phys. Lett. **76**, 2588 (2000).

[12]  B. Van Waeyenberge, A. Puzic, H. Stoll, K. W. Chou, T. Tyliszczak, R. Hertel, M. Fahnle, H. Bruckl, K. Rott, G. Reiss, I. Neudecker, D. Weiss, C. H. Back, and G. Schutz, Nature **444**, 461 (2006).

[13] R. K. Dumas, C.-P. Li, I. V. Roshchin, I. K. Schuller, and K. Liu, Phys. Rev. B **75**, 134405 (2007).

[14] R. K. Dumas, K. Liu, C. P. Li, I. V. Roshchin, and I. K. Schuller, Appl. Phys. Lett. **91**, 202501 (2007).

[15] C. R. Pike, C. A. Ross, R. T. Scalettar, and G. T. Zimanyi, Phys. Rev. B **71**, 134407 (2005).

[16] J. E. Davies, O. Hellwig, E. E. Fullerton, G. Denbeaux, J. B. Kortright, and K. Liu, Phys. Rev. B **70**, 224434 (2004).

[17] M. Chen, C. L. Chien, and P. C. Searson, Chem. Mater. **18**, 1595 (2006).

[18] J. D. Medina, M. Darques, T. Blon, L. Piraux, and A. Encinas, Phys. Rev. B **77**, 014417 (2008).

[19] V. Rose, X. M. Cheng, D. J. Keavney, J. W. Freeland, K. S. Buchanan, B. Ilic, and V. Metlushko, Appl. Phys. Lett. **91**, 132501 (2007).

[20] S. J. Cho, A. M. Shahin, G. J. Long, J. E. Davies, K. Liu, F. Grandjean, and S. M. Kauzlarich, Chem. Mater. **18**, 960 (2006).

[21] M. Donahue and D. Porter, "OOMMF User's Guide, Version 1.0", Interagency Report NISTIR, 6376 (1999). Material parameters suitable for polycrystalline Co dot were used (saturation magnetization $M_S = 1.4 \times 10^6$ A/m, exchange stiffness $A = 3 \times 10^{-11}$ J/m, anisotropy was neglected, cell size = 2 nm).

[22] T. R. McGuire and R. I. Potter, IEEE Trans. Magn. **11**, 1018 (1975).

**Figure captions:**

Fig. 1. (Color online) Room temperature magnetic hysteresis loops for 200 nm diameter nanowires of (a) [Co(10nm)/Cu(250nm)]$_5$ and (b) [Co(55nm)/Cu(350nm)]$_4$. A loop shape analysis for nanowires with various Co thicknesses (Cu thickness is 250-350nm) is shown in (c), where the loop width $W$ at various $M/M_S$ values is compared with that at zero magnetization. Lines are guides to the eye.

Fig. 2. (Color online) Families of room temperature first order reversal curves (FORC's) for 200nm diameter (a) [Co(10nm)/Cu(250nm)]$_5$ and (c) [Co(55nm)/Cu(350nm)]$_4$ multilayered nanowires measured with the magnetic field perpendicular to the wires. Black dots represent the starting points of each FORC. The corresponding FORC distributions are shown in (b) and (d) respectively.

Fig. 3. (Color online) Magnetoresistance of 200nm diameter [Co(50nm)/Cu(10nm)]$_{150}$ multilayered nanowires measured at 300 K with the applied field (a) parallel and (b) perpendicular to the wires. Decreasing and increasing field sweeps are represented by solid and open circles, respectively.

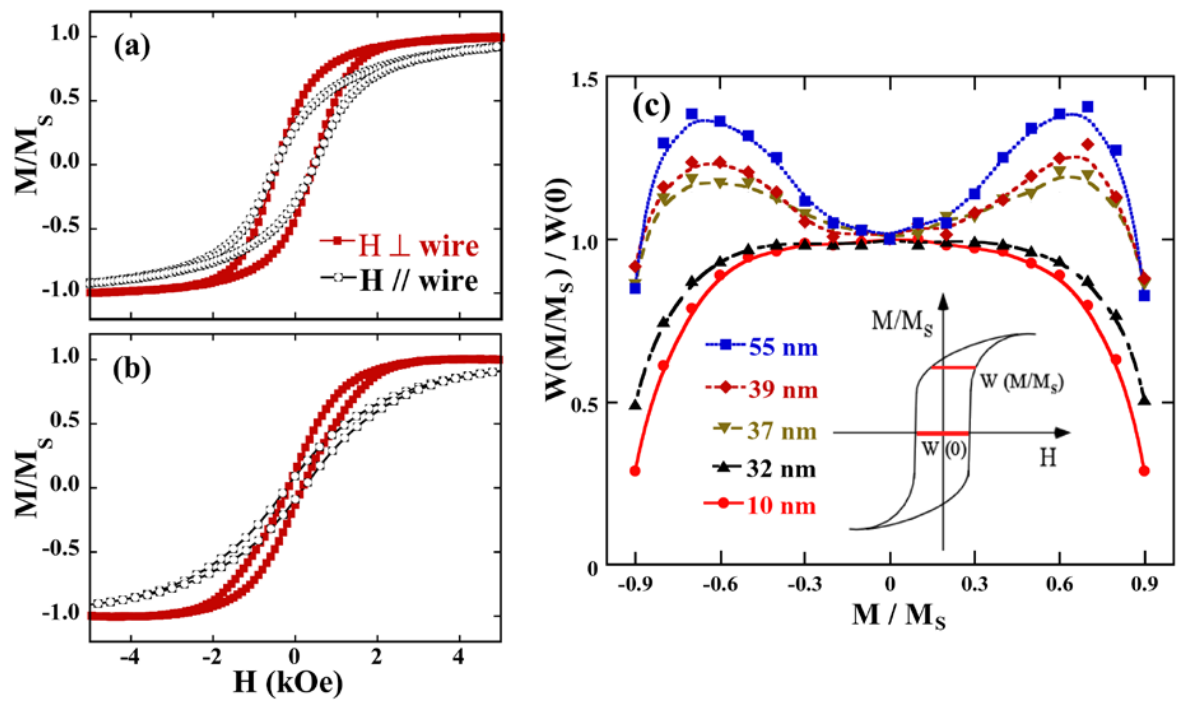

**Fig. 1.**

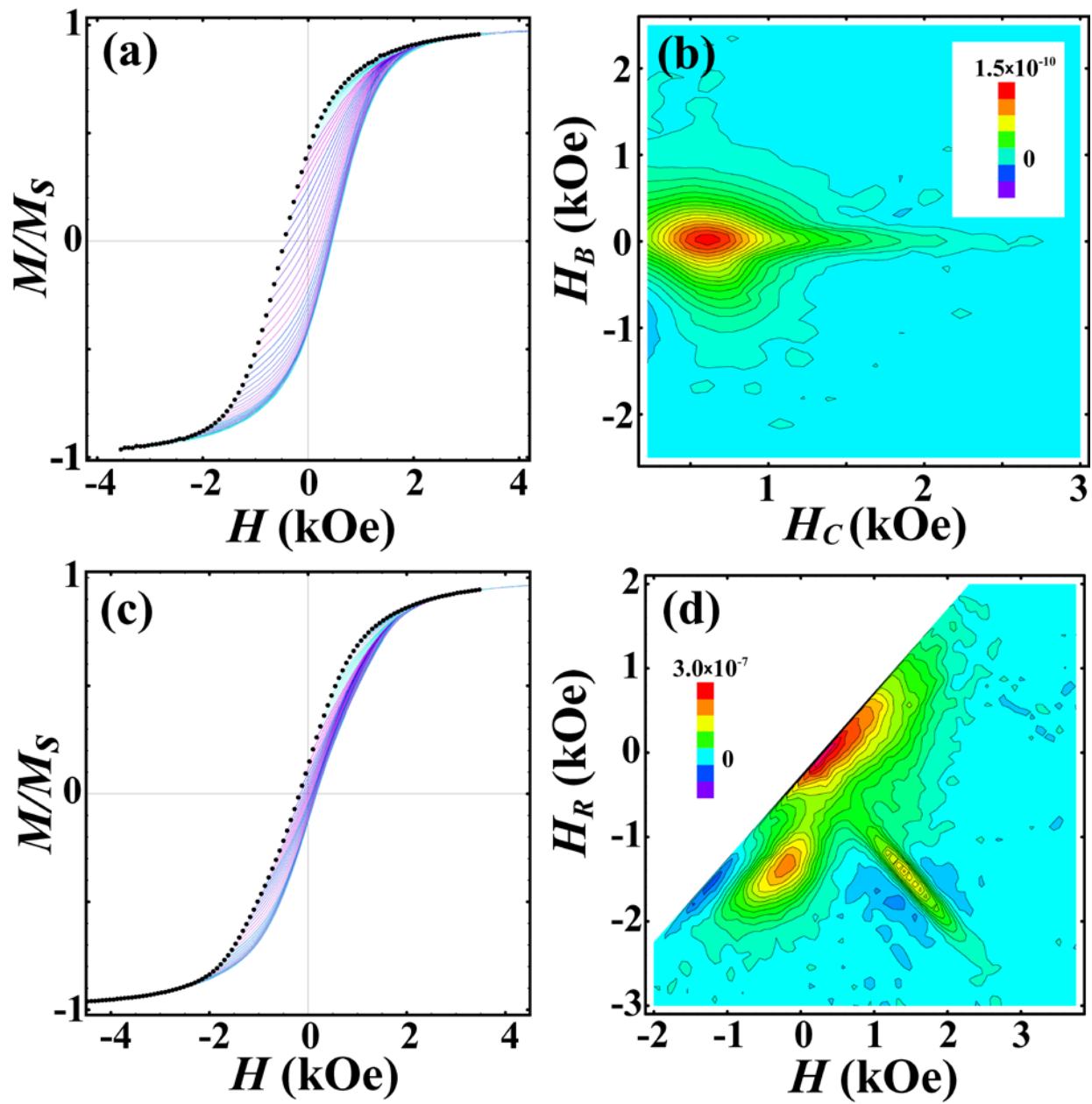

**Fig. 2.**

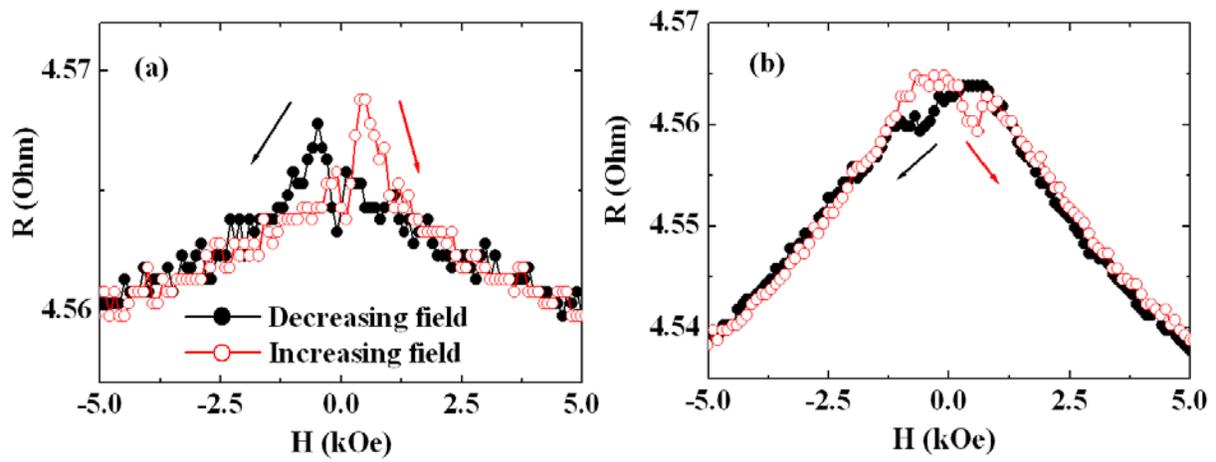

Fig. 3.